\documentclass[runningheads]{llncs}

\usepackage{amsmath,amssymb}
\usepackage{hyperref}
\usepackage{academicons}
\usepackage{xcolor}
\usepackage{cite}
\usepackage{multirow}

\usepackage{graphicx}
\newcommand{\ket}[1]{\left|#1\right\rangle}

\newcommand{\orcid}[1]{\href{https://orcid.org/#1}{\textcolor[HTML]{A6CE39}{\aiOrcid}}}

\begin{document}
\title{Simulation Methodology for Electron Transfer in CMOS Quantum Dots
}
\titlerunning{Simulations Methodology for Transport in CMOS Quantum Dots}
%

\author{
Andrii Sokolov\inst{1}
\and
Dmytro Mishagli\inst{1}
\and
Panagiotis Giounanlis\inst{1}
\and
Imran Bashir\inst{2}
\and
Dirk Leipold\inst{2}
\and
Eugene Koskin\inst{1}
\and
R. Bogdan Staszewski\inst{1,2}
\and
Elena Blokhina\inst{1}
}

\authorrunning{A. Sokolov et al.}
%
\institute{
University College Dublin, Belfield, Dublin 4, Ireland 
\and
Equal1 Labs, Fremont, CA 94536, USA}

\maketitle              
%

\begin{abstract}
The construction of quantum computer simulators requires advanced software which can capture the most significant characteristics of the quantum behavior and  quantum states of qubits in such systems. Additionally, one needs to provide valid models for the description of the interface between classical circuitry and quantum core hardware. 
In this study, we model electron transport in semiconductor qubits based on an advanced CMOS technology. Starting from 3D simulations, we demonstrate an order reduction and the steps necessary to obtain ordinary differential equations on probability amplitudes in a multi-particle system. We compare numerical and semi-analytical techniques concluding this paper by examining two case studies: the electron transfer through multiple quantum dots and the construction of a Hadamard gate simulated using a numerical method to solve the time-dependent Schr\"odinger equation and the tight-binding formalism for a time-dependent Hamiltonian.

\keywords{CMOS quantum dots \and charge qubits \and position-based charge qubits \and tight binding formalism \and split-operator method \and electron transfer \and X-rotation gate}
\end{abstract}

\section{Introduction}

This work is motivated by the development of Complementary Metal-Oxide-Semiconductor (CMOS) charge qubits in one of the most advanced technologies, 22FDX by
GlobalFoundries employing a 22~nm Fully-Depleted Silicon-On-Insulator (FDSOI) process. Building charge, spin or hybrid qubits by exploiting the fine-feature lithography of CMOS devices is currently a dominant  trend towards a large-scale quantum computer~\cite{Bashir_2019,Yang_2019,Fujisawa_2004,Sarma_2011,Li_2018,Veldhorst_2015,Fujisawa2}. Although of all the mentioned CMOS silicon qubits, charge qubits have quite short decoherence time~\cite{Weichselbaum_2004,petersson2010quantum}, they are revisited now in light of the high material interface purity, fine feature size and fast speed of operation of the latest mainstream nanometer-scale CMOS process technology (see the review section in Ref.~\cite{Blokhina_2019}). The speed of operation of quantum gates can now be ultra short due to the transistor cut-off frequency reaching half-terahertz in advanced CMOS. Thus, the number of quantum gate operations can be on the same order as with spin-based or hybrid qubits. For this reason, the paper is focused on the modelling of charge qubits.

The development of quantum computer emulators requires high-level software representing the physics of quantum structures under study with high accuracy. At the same time, it should incorporate the control of quantum states of a large number of qubits, the realization of quantum gates, the act of measurement and transport of quantum information including the interface between classical circuitry and quantum registers, mentioning some of the major components needed for such an attempt. To facilitate an accurate modeling of interacting qubits, one is required to solve a many-body quantum system, which is a very resource intense task for classical simulators. There are some methods from computational quantum physics that could be used as a first approach for such problems. The density function theory (DFT) is one of the most known and, in principle, exact \emph{ab initio} method~\cite{remediakis1999band, zheng2018real}. However, it is not applicable in practice for strongly correlated electron problems, since the exact exchange correlation functional is not known. In principle, the electrons in the position-based charge qubits described in this study are strongly interacting by Coulomb interaction \cite{szafran20}. From this point of view, any single-electron method such as DFT cannot be reliably used. Configuration interaction methods from quantum chemistry may also be considered exact \emph{ab initio} methods, but of course they are limited to extremely small system sizes due to the well-known exponential scaling of the Hilbert space with system size, making such methods essentially useless for our present purposes. By contrast, virtually all methods in quantum physics employ effective (or reduced) models that capture the essential physics of interest while throwing away irrelevant details~\cite{cubitt2018universal}. Such approaches can even be quantitative semi-empirical, by fitting model parameters to experimental data.

In this study, we will focus on the modelling of a specific type of CMOS silicon qubits known as position-based charge qubits (or simply as charge qubits)~\cite{Blokhina_2019}.
Starting from 3D simulations, we demonstrate an order reduction and the steps necessary to obtain ordinary differential equations on probability amplitudes in a multi-particle system. We compare numerical and semi-analytical techniques concluding this paper by examining two case studies: electron transfer through multiple quantum dots and construction of a Hadamard gate simulated using the time-dependent Schr\"odinger equation and the tight-binding formalism. 


\section{Coupled Quantum Dot Chains and Structures in 3D}\label{sec:3D}

\begin{figure}[t]
\center
\includegraphics[width=0.80\textwidth]{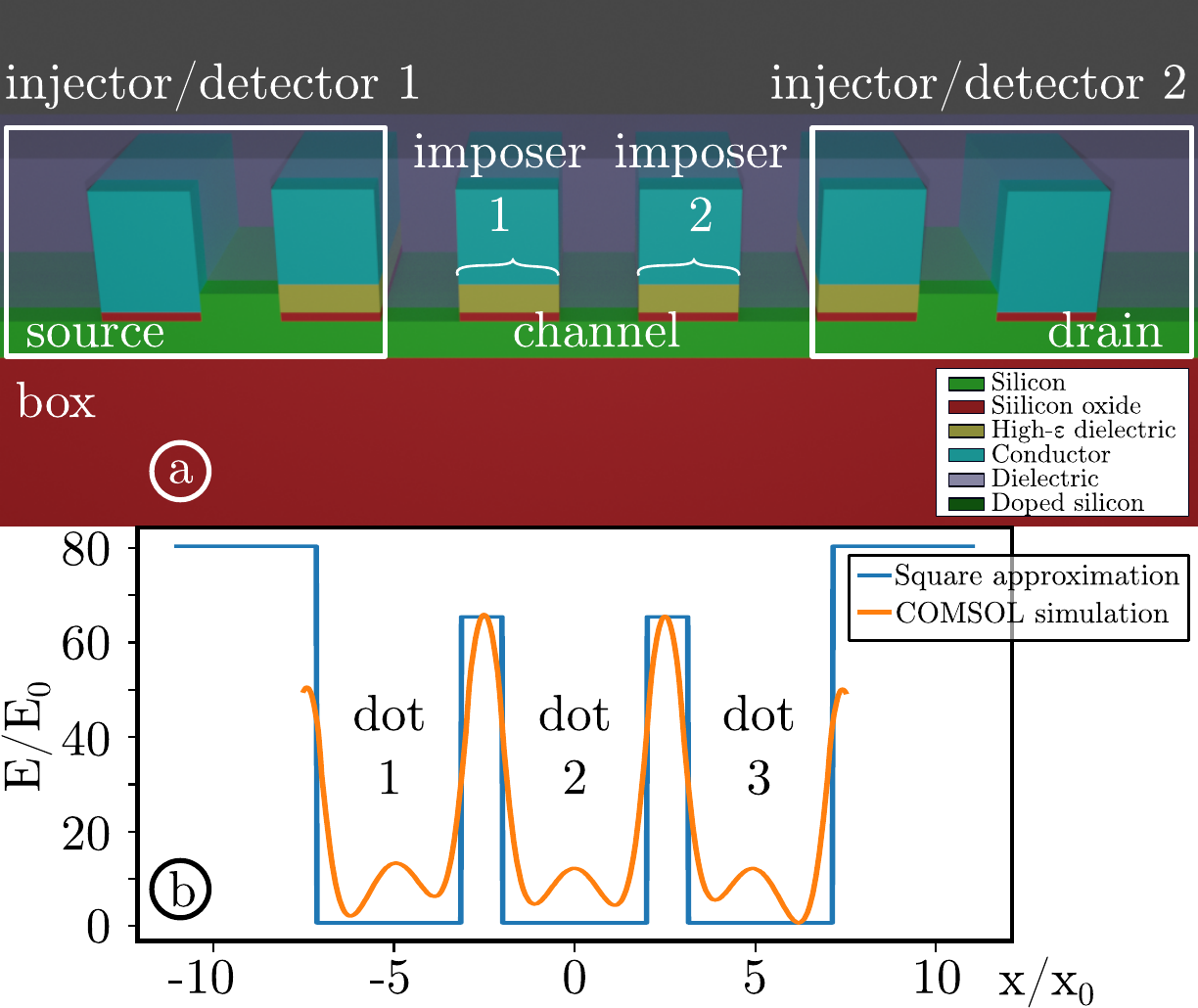}
\caption{(a) Schematic diagram of a CMOS structure implementing three QDs including two imposers between the QDs and two injectors/detectors at the edges. (b) Normalised potential energy as a function of the position obtained from FEM electromagnetic simulations (orange line) and a simplified piece-wise potential energy function (blue line).}
\label{fig:potential_profile}
\end{figure}

In this section, we provide the general description of the structures containing coupled quantum dots (QDs) under study and key results of their Fine-Element Method (FEM) modelling employing COMSOL Multiphysics. As we aim to penetrate the behaviour of a quantum processor fabricated in a commercial technology~\cite{Bashir_2019}, FEM simulations have been carried out on the structures whose dimensions and composition are inspired by the 22FDX technology of GlobalFoundries. The schematic 3D structure under study is shown in Fig.~\ref{fig:potential_profile}(a). The parameters and normalisation units used are presented in Table~\ref{tab:dim}. Each structure can be seen as  transistor-like devices arranged in a chain. Each `transistor' contains a control gate (which we call an imposer) made of a very thin SiO$_2$ layer, high-$k$ dielectric layer and a thick heavily doped polysilicon layer. Beneath a gate, there is a thin depleted silicon channel and a thin buried oxide (BoX) layer deposited on a thick lightly doped silicon wafer. An insulating coat is usually deposited on the top of the entire structure for electric isolation. The edges of the structure are connected to classic circuit devices known as single-electron injectors (detectors).

\begin{table}[t!]
\center
\caption{Parameters used for simulations in this study}\label{tab:dim} 
\begin{tabular}{|l|ll|}
\hline
Elementary charge, $e$ & $1.602\times10^{-19}$& C \\
\hline
Effective mass, $m_e^*$ & $1.08\times9.109\times10^{-31}$ & kg \\
 \hline
Length unit, $x_0$ & $20$ & nm \\
 \hline
\multirow{2}{*}{Energy unit, $E_0$} &  $\hbar^2/2m_e^*x_0^2= 1.41\times10^{-23}$& J \\
& $=87.6$ & $\mu$eV \\
 \hline
Time unit, $t_0$ & $2\pi\hbar/E_0 = 47.3$ & ps \\
 \hline

\end{tabular}
\end{table}


Semiconductor equilibrium simulations using the Semiconductor module of COMSOL Multiphysics have been carried out for temperatures ranging from 10 to 100~K to confirm that the silicon channel is completely depleted and there are no thermally generated carriers or carriers diffused from the interconnections with the heavily doped regions belonging to the injectors and detectors \cite{Bashir_2019}. Electromagnetic simulations have been carried out to investigate the electric potential and electric field propagation through the structure when a given voltage is applied at the imposers. The electric potential developed in the silicon channel allows us to calculate the modulation of the edge of the conduction band and the formation of potential energy wells in the channel. An  example of the potential energy on the surface of the silicon channel as a function of the coordinate along the channel is shown in Fig.~\ref{fig:potential_profile}(b) where one can observe three potential wells forming in the region between the imposers. 

In general, there exist specific imposer voltages so that potential energy ``barriers'' are formed under the imposers and ``wells'' are formed in between the imposers. We refer to the region between the imposers where a potential energy well is formed as a quantum dot.
In the figure, the minimum of the potential energy is conventionally placed at 0~meV. In a typical scenario, barriers of 1 to 4~meV can be formed when a sub-threshold voltage is applied at the imposers.
The resulting potential energy can be effectively approximated by an equivalent piece-wise linear function. The time-independent Schr\"odinger equation (TISE) with a piece-wise linear potential energy can be solved analytically or numerically in one, two or three dimensions to find a set of eigenfunctions and eigenenergies. The time-dependent Schr\"odinger equation (TDSE) can also be solved (as we will show later in the paper), but even in the 1D case, it requires substantial computational resources. The next section outlines the problem overview and the assumptions taken in this study.

\section{Outline of Quantum Mechanical Modelling}\label{sec:outline}

As the main model, we consider an isolated quantum system without external fluctuations. However, we will also present some results including effects of decoherence, which arise from the coupling between the quantum system under consideration with the environment. In this paper, the coupling appears as time-dependent noise terms in the effective Hamiltonian, but it should be noted that there are a number of approaches to model decoherence effects. Coupling to a fermionic bath leads to Kondo-type physics and the low-energy quenching of spin qubits, while coupling to a bosonic environment, leads to the localization through the Caldeira-Legget mechanism~\cite{mitchell2010two}. In general, we note that both kinds of processes occur at some characteristic timescale (not instantaneously).  In addition, we neglect other degrees of freedom (spin, valley and orbital) since we are interested in electron transfer in charge qubits (also called position-based qubits, as stated earlier). 

We start from the fundamental equation in Quantum Mechanics, the Schr\"o\-din\-ger equation.
We have made some assumptions, which reduce the complexity of the problem:
\begin{enumerate}
\item We are interested in the transfer of an electron through the silicon channel of a quantum register containing multiple QDs. Since the channel is very thin, the electron wave function is ``shallow'' and we can neglect one dimension. In addition, FEM simulations show the symmetry of the wave function, hence we use the reduction of the wave function along the symmetry line. In this case,  it is reasonable to use the 1D modelling~\cite{Giounanlis_2019}.
\item The complexity of the multi-electron model is significantly higher than that of a single-electron one. Our first step is to consider the transfer of one electron through the channel by changing the imposer voltages.  
\item In general, any changes of imposer voltages may cause  non-trivial changes to  the potential energy  which can cause second-order effects on the wave function evolution. At the moment we do not take into account second- and third-order effects associated with abrupt imposer voltage changes of electronic noise.
\end{enumerate}

The starting point is the 1D time-dependent Schr\"odinger equation:
\begin{equation}
 i\hbar\frac{\partial \Phi(x,t)}{\partial t} = \left[-\frac{\hbar^2}{2 m_e^*}\frac{\partial^2}{\partial x^2} + V(x,t) \right] \Phi(x,t),
\label{eqn:sch}
\end{equation}
where $\hbar$ is the reduced Plank's constant, $m_e^*$ is the effective mass of the electron, $\Phi(x,t)$ is the wave function of the particle, and $V(x,t)$ is the 1D potential energy function. The the potential energy of the electron is obtained from the simulations presented in Sec.~\ref{sec:3D}, see Fig.~\ref{fig:potential_profile}. The potential energy calculated from the FEM simulations and its changes due to the variation of imposer voltages are exported as a table function from the COMSOL electrostatic simulator and is used as a `pre-generated' input parameter of the quantum simulator.

In order to have a useful simulator of the studied quantum structure, one needs to define a  possible ``localised'' state of an electron injected in the structure, simulate its evolution with time at a given potential energy along the structure and calculate the probability of the electron to be measured at the edges of the structure by a detector device. The eigenenergies and eigenfunctions of the time independent case are particularly useful since they define the frequency of transitions, and, as the next step, we address the calculation eigenfunctions for an arbitrary shaped potential energy.

\section{Eigenenergy and Eigenfunction Calculation Using the Matrix Diagonalisation Method}\label{sec:matrix}

In this section, we discuss a method for obtaining an approximate solution to a one-dimensional TISE~\cite{marsiglio09}. A distinguishable feature of this method is that it allows one to obtain all the bound states and their corresponding wave functions at once. The method is closely related but not equivalent to the perturbation theory~\cite{b_landafshitz_quant1981}. Assume that an electron is confined by a finite potential $V(x)$ and this finite potential is located \emph{inside} an infinite potential well $V_{\infty}(x)$, so that the Hamiltonian $\hat H$ of such a system reads
\begin{equation}\label{eq:schr_operator}
	\hat H = \hat H_0 + \hat V.
\end{equation}
Here $\hat H_0$ is the Hamiltonian operator that corresponds to the problem of a particle (an electron) in an infinite potential well, $\hat H_0 = - \hbar^2/2m \cdot d^2/dx^2 + V_{\infty}(x)$. The eigenvalues $E_n^{(0)}$ and eigenfunctions $\psi_n^{(0)}$ of such an operator are well known. The eigenfunctions $\psi_n$ and eigenvalues $E_n$ of the operator \eqref{eq:schr_operator} are the solutions to the equation
\begin{equation}\label{eq:tise}
	(\hat H_0 + \hat V) \psi_n = E_n\psi_n.
\end{equation}
Since the functions $\psi_n^{(0)}$ form a complete basis set, for the required functions $\psi_n$ we can write the expansion
\begin{equation}\label{eq:psi_expansion}
	\psi_n = \sum\limits_m c_{nm} \psi_m^{(0)},
\end{equation}
where $c_{nm}$ are unknown coefficients. Thus, the initial equation \eqref{eq:tise} yields
\begin{equation}
	\sum\limits_m c_{nm}(\hat H_0 + \hat V) \psi_m^{(0)} = E_n \sum\limits_m c_{nm} \psi_m^{(0)}.
\end{equation}
Multiplying both sides of the latter equation by $\psi_k^{(0)*}$ and integrating over the variable $x$, we obtain

\begin{equation}\label{eq:tise_matrix}
	\sum\limits_m c_{nm} H_{km} = E_n c_{nk}, \qquad \text{where} \qquad H_{km} = E_k^{(0)} \delta_{km} + V_{km}.
\end{equation}
The matrix element $V_{km}$ is determined by the potential $V(x)$:
\begin{align}
	V_{km} &= \int \psi_k^{(0)*}V(x)\psi_m^{(0)}dx.\nonumber
	\\
	&= \frac{2}{L} \int\limits_0^L \sin\left( \frac{k\pi x}{L} \right ) V(x) \sin\left( \frac{m\pi x}{L} \right ) dx
	\label{eq:potential_matrix}
\end{align}
Thus, the problem of solving the differential equation \eqref{eq:tise} is reduced to the square matrix equation problem \eqref{eq:tise_matrix} that involves the integrals \eqref{eq:potential_matrix}. The coefficients $c_{nm}$ form the eigenvectors of the Hamiltonian matrix $H_{km}$ where $n$ stays for the index of a given bound energy level. Such a problem is usually solved by a matrix diagonalisation procedure, hence the name of the method. Below we discuss different potentials $V(x)$ and their impact on the solution.


In case of a square well (potential barrier) structure, the integration in \eqref{eq:potential_matrix} is trivial. Indeed, for a piece-wise potential that takes either zero or some constant value $V_i$ for $x \in [a,b]$, we have
\begin{equation}
	V_{km} = V_i \times
	\begin{cases}\displaystyle
	\frac12 \left. \left( x - \frac{L}{\pi k^2}\sin\frac{k^2\pi}{L}x \right) \right|_a^b, & k=m;
	\\
	\displaystyle
	\frac{L}{2\pi} \left. \left( \frac{\sin (k-m)\pi x/L}{k-m} + \frac{\sin (k+m)\pi x/L}{k+m} \right) \right|_a^b, & k \neq m.
	\end{cases}
\end{equation}

When the potential $V(x)$ is given by some continuous but known function, the integration in \eqref{eq:potential_matrix} should not be problematic. While it can be tedious, in principle this is a solvable problem. 

\begin{figure}[t!]
\center
\includegraphics[width=\textwidth]{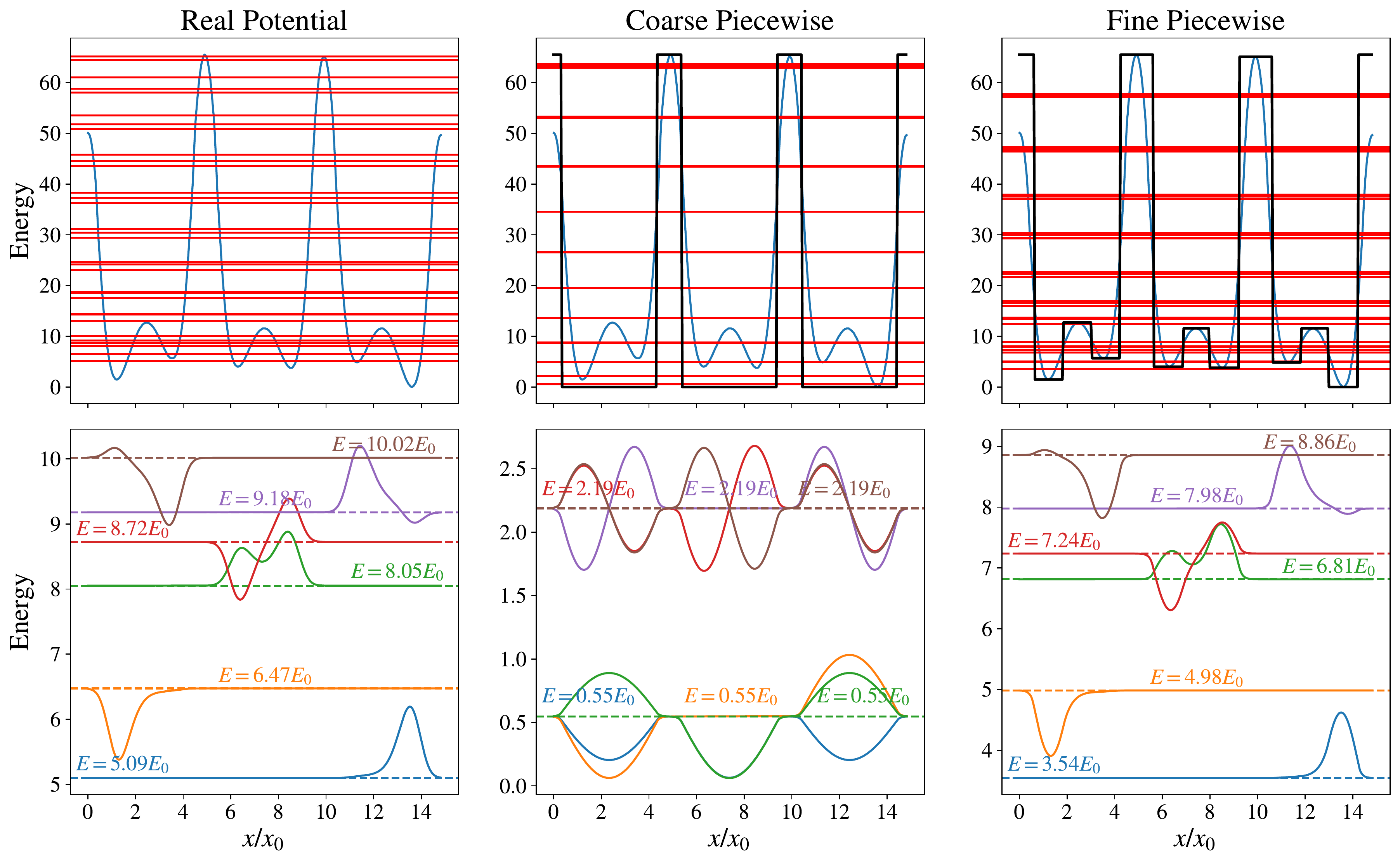}
\caption{Energy spectra of the potential generated from FEM simulations (COMSOL) and the six lowest energy levels with the corresponding wave functions for the three cases: realistic potential function integrated with splines (left), coarse piece-wise approximation (middle) and fine piece-wise approximation (right).
\textit{Left, realistic potential}: the lowest energy levels (as well as the corresponding wave functions) exhibiting  bound states `localised' in each well.
\textit{Middle, coarse piece-wise approximation}:  the lowest energy levels are grouped in three, exhibiting `non-localised' bound states.
\textit{Right, fine piece-wise approximation}: the lowest energy levels are of the same order as those of the realistic potential energy; the wave functions match qualitatively.
}
\label{fig:comsol_potential_inside_inf_well}
\end{figure}

In the present study, we use the smooth potential energy generated by FEM simulations as shown in Fig.~\ref{fig:potential_profile}. For comparison, we create its coarse piece-wise approximation and also its fine piece-wise approximations as shown in Fig.~\ref{fig:comsol_potential_inside_inf_well}. 
The realistic smooth potential energy function is given in form of data points, and thus the easiest approach to the calculation of \eqref{eq:potential_matrix} is numerical integration (e.g., using splines). Placing the potential inside an infinite well of length $L$ and separating it from the edges with a distance $h$, as described above, we obtain equivalent formulation of the problem within the Matrix Diagonalisation Method.

The result of the eigenenergy and eigenfunction calculations for the realistic potential energy\footnote{The source code for this section is available from www.github.com/mishagli/qsol} is shown in Fig.~\ref{fig:comsol_potential_inside_inf_well} in the left column. We have used 320 basis functions, and the width of the surrounding infinite well was $L=44.472$ of dimensionless units, which gave 32 bound states. One can see that such a potential has a nontrivial asymmetric allocation of the wave functions. The coarse and fine piece-wise approximations to the realistic function are shown in the same figure in the middle and right columns correspondingly. In the case of the coarse approximation, when the fine features of the potential energy  such as ``double-wells''are ignored, the lowest energy levels are grouped so that they are hardly distinguishable. There is also a significant difference in the values of the eigenenergies compared to the realistic case. On the other hand, the fine piece-wise approximation allows one to preserve both qualitative and quantitative properties of the real potential energy. 

Solving the time independent problem gives a good understanding of the system's properties at  early stages of modelling. The time-dependent solution is discussed next.

\section{Split-Operator Method}\label{sec:split}

Since we would like to model the evolution of an electron injected into a register of coupled QDs and  subject to a variable potential energy, we employ a Split-Operator Method (SOM) to solve the time-dependent problem~\cite{bandrauk1991improved,hermann1988split}. This method has been chosen due to its effectiveness at a moderate computational cost. Since the method is well documented, we provide only its brief description, discuss its computational cost and show its application to the studied system.

\begin{figure}[t]
\center
\includegraphics[width=1.0\textwidth]{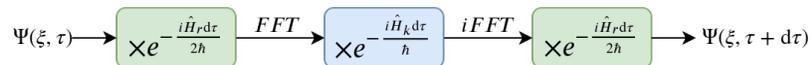}
\caption{High-level algorithm illustrating the steps of the split-operator method.} \label{fig:split_op}
\end{figure}

The split-operator method is based on the fact that in conventional cases it is possible to split the Hamiltonian operator into two components, one being position dependent $\hat{H}_r$ and the other being momentum dependent $\hat{H}_k$~ \cite{pahlavani2012theoretical}. Moreover, it is possible to show that in the position space, the $\hat{H}_r$ operator is reduced to the multiplication of the wave function by a position dependent function. The same reduction can be done for the momentum dependent component of the Hamiltonian operator in the momentum space. This concept is illustrated in Fig.~\ref{fig:split_op}. 

This method allows one to observe the evolution of the wave function with time, which is particularly useful in the case of a time-dependent Hamiltonian. It also allows one to calculate the ground state of a system (in this case one can use the imaginary time step $i\mathrm{d} t$). In our case, when we deal with multiple potential wells (or as we say coupled quantum dots), it allows us to obtain corresponding localised states. The construction of the localised states in a system with the potential energy from Section~\ref{sec:3D} by the use of SOM is compared with that obtained by applying a unitary transformation on the eigenfunction set (obtained in Sec.~\ref{sec:matrix}). Both methods show a good agreement. However, it is more convenient to use the matrix diagonalisation method in order to obtain the full spectrum of possible states.

With regard to the computational cost of this method, it has two contributing aspects: 
\begin{itemize}
\item [$\square$] The calculation of the forward and inverse Fast Fourier Transforms (FFT and iFFT correspondingly) is proportional to the number of discretisation points $n$  as $n \ln(n)$. A higher number $n$ provides a higher resolution for the wave function, and therefore for the probability density function. Note that in order to use the FFT algorithm, the number of coordinate points should be a power of 2 (in our case, 512).
\item [$\square$] This operation is repeated over a desired number of  time steps $\mathrm{d} \tau$. The accuracy of the SOM  is $O(\mathrm{d}t^3)$, and thus the time step operation should be done with a relatively small time step (in our case $10^{-4}$ of dimensionless units of time). 
\end{itemize}

After we have implemented the split-operator method and ensured that it gives consistent results, our next task is to simulate the behaviour of an electron in a quantum register. We will show two examples: the transport of an electron from the first to the last QD in the register combined with electron probability oscillations and ``spliting'' electron's wave function between a number of QDs. These simulations are presented in a later section using the coarse piece-wise approximation and by adjusting the height of the potential barriers separating the QDs. 

\section{Multiple Quantum Dot Model in the Tight Biding Formalism}
\label{sec:tb_model}

In this section, we will introduce the tight binding formalism. This will allow one to capture the time dependent dynamics assuming an ideal quantum transport in the quantum structure (register) and can be easily extended to multi-particle and multi-energy level  systems~\cite{Blokhina_2019,Giounanlis_2019mdpi}. We use this approach for additional verification of the SOM modelling. 

In the tight binding formalism, we assume that electrons can be represented by wave functions associated with localised states in a discrete lattice. We can visualize the quantum register of Fig.~\ref{fig:potential_profile}(a) as a pseudo-1D lattice. For each QD, we consider one effective quantum state, which can be represented as $\ket{j}$, where $j=1,2,3...$ is an integer denoting the position of each QD. Then, the Hamiltonian of the system can be written as:
\begin{equation}
\mathbf {\hat H}= -t_s \left(\sum_{j k} \hat{c}_{j}^{\dagger} \hat{c}_{k} + \hat{c}_{k}^{\dagger} \hat{c}_{j}\right)
\end{equation}
where $\hat{c}_{j}^{\dagger}\left(\hat{c}_{j}\right)$ are the creation (annihilation) operators, which create (annihilate) a particle at site~$j$. The terms $t_{s,jk}$ are describing  tunnelling, i.e., the hopping of electrons from site~$j$ to site~$k$ in the lattice. Here, we will consider the symmetrical case, and therefore we will omit the potential energy of each QD (since it only shifts the total energy of the system by a constant and has no physical contribution to its evolution). We will also assume that sites $k$ and $j$ are immediate neighbors, and we will disregard any probability of a particle hopping to a distant QD. It should be noted that for single-electron tunneling processes through potential barriers, the tunneling rate is exponentially suppressed in the barrier height and width, therefore this is a reasonable assumption. In this representation, the wave function can be expressed as a linear combination in the position basis:
%
\begin{equation}
\ket{\Psi} = \sum_k c_k(t) \ket{k}
\end{equation}
where $c_k(t)$ are the complex probability amplitudes of the states $\ket{k}$, and the normalisation restriction applies $\sum_k |c_k(t)|^2$ $= 1$.
Furthermore, by considering the time-dependent Schr\"odinger equation~\cite{Blokhina_2019}, one can write:
\vspace{-1mm}
\begin{equation}
i \hbar \frac{d \textbf{C}(t) }{d t} = \hat {\text H}(t) \textbf{C}
\label{eq:SE}
\end{equation}
where $\textbf{C}(t) = \{c_1(t), c_2(t), \ldots, c_k(t)\}$ is the vector of the probability amplitudes.

In this work, we are interested in describing the dynamics of the system for a time-dependent case where the hopping coefficients $t_{s,jk}$ do not remain constant with time (and can be increased or decreased by means of controlling the potential energy function). In order to perform quantum operations, one needs to apply a correct sequence of voltage pulses at the imposers (gates) at specific time instances, changing the tunnelling probabilities. In such a case, the Hamiltonian of the system is changing with time.
In the system under study we can consider a sudden change in the Hamiltonian~\cite{song2015condition}. Also, the applied external fields (square voltage pulses) will be assumed of small magnitudes. As a consequence, the Hamiltonian of the system takes the form:
\begin{equation}
\mathbf{\hat H}(t) = \begin{bmatrix}
  0 & t_{h,21}(t) & 0 \\
  t_{h,12}(t) & 0 & t_{s,32}(t)\\
  0 & t_{s,23}(t) & 0 
  \end{bmatrix}
\end{equation}
where each of the $t_{h,jk}$ hopping terms is a piece-wise time dependent-function of the form:

\begin{equation}
    t_{h,jk}=\left\{\begin{matrix}
    t_{h,\text{low}} & t < t_{jk,0} \\
    t_{h,\text{high}} & t_{jk,0} \leq t_{jk} \leq t_{jk,0} + t_{jk,\text{width}}\\
    t_{h,\text{low}} & t > t_{jk,0} + t_{jk,\text{width}}.
    \end{matrix}\right.
\end{equation}
where $t_{jk,0}$ is the initial time instance when the pulse is applied at the imposer separating the quantum dots $j$ and $k$, and $t_{jk,\text{width}}$ is the time duration for which the pulse is applied. Note that one needs to match the initial conditions between the solutions of~\eqref{eq:SE}, before and after the application of the pulse. The values used in this simulation are: $t_{h,\text{high}} = 2.82 E_0$, $t_{h,\text{low}} = 0.0013 E_0$. 

\begin{figure}[t!]
\center
\includegraphics[width=0.8\textwidth]{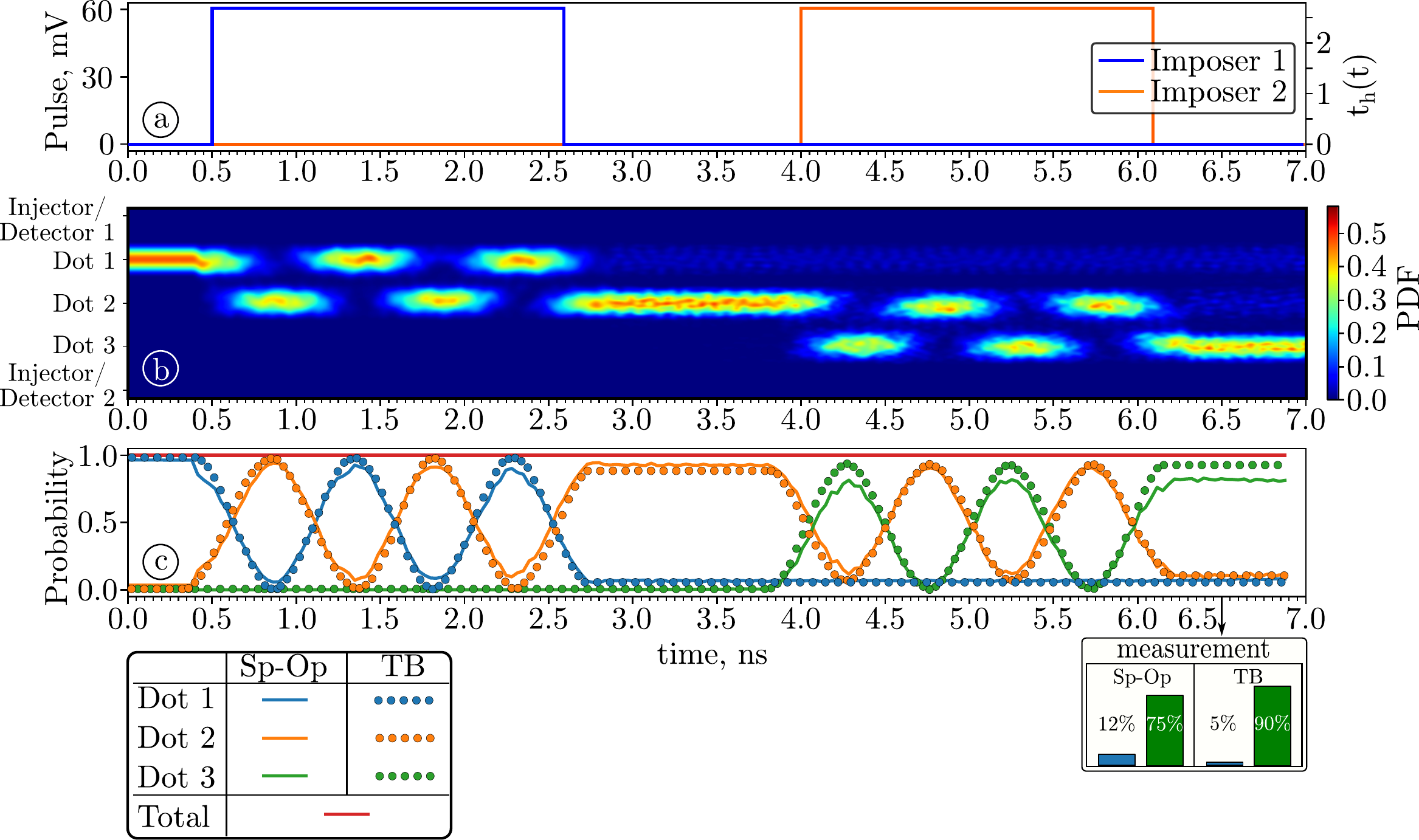}
\caption{Transport of a single injected electron from dot 1 to dot 3. (a) Control pulses applied to imposers and corresponding hopping terms. (b) Probability heatmap generated by the SOM showing electron transfer and Rabi oscillations between pairs of dots. (c) Probability of detecting an electron in each quantum dot (lines correspond to the SOM, circles~--- to the tide-binding formalism).} \label{fig:split_transport}
\end{figure}

\begin{figure}[t!]
\center
\includegraphics[width=0.8\textwidth]{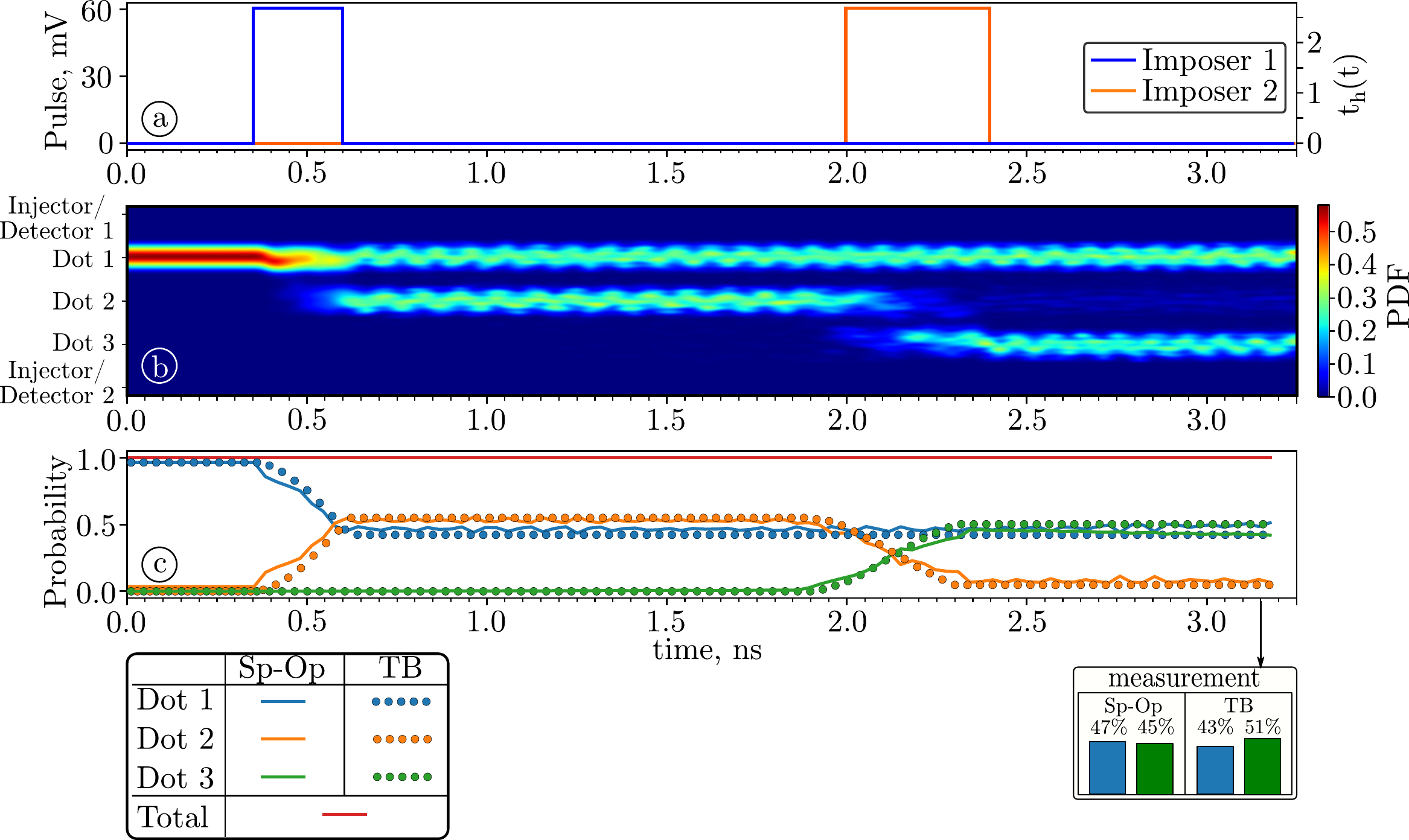}
\caption{Rotation gate operation applied onto a single injected electron in dot 1 and detected at both detectors. (a) Control pulses applied to imposers and corresponding hopping terms. (b) Probability heatmap generated using the SOM showing the reaction of the electron to the pulses. (c) Probability of detecting the electron in each quantum dot.} \label{fig:split_hadamard}
\end{figure}

\begin{figure}[t!]
\center
\includegraphics[width=0.8\textwidth]{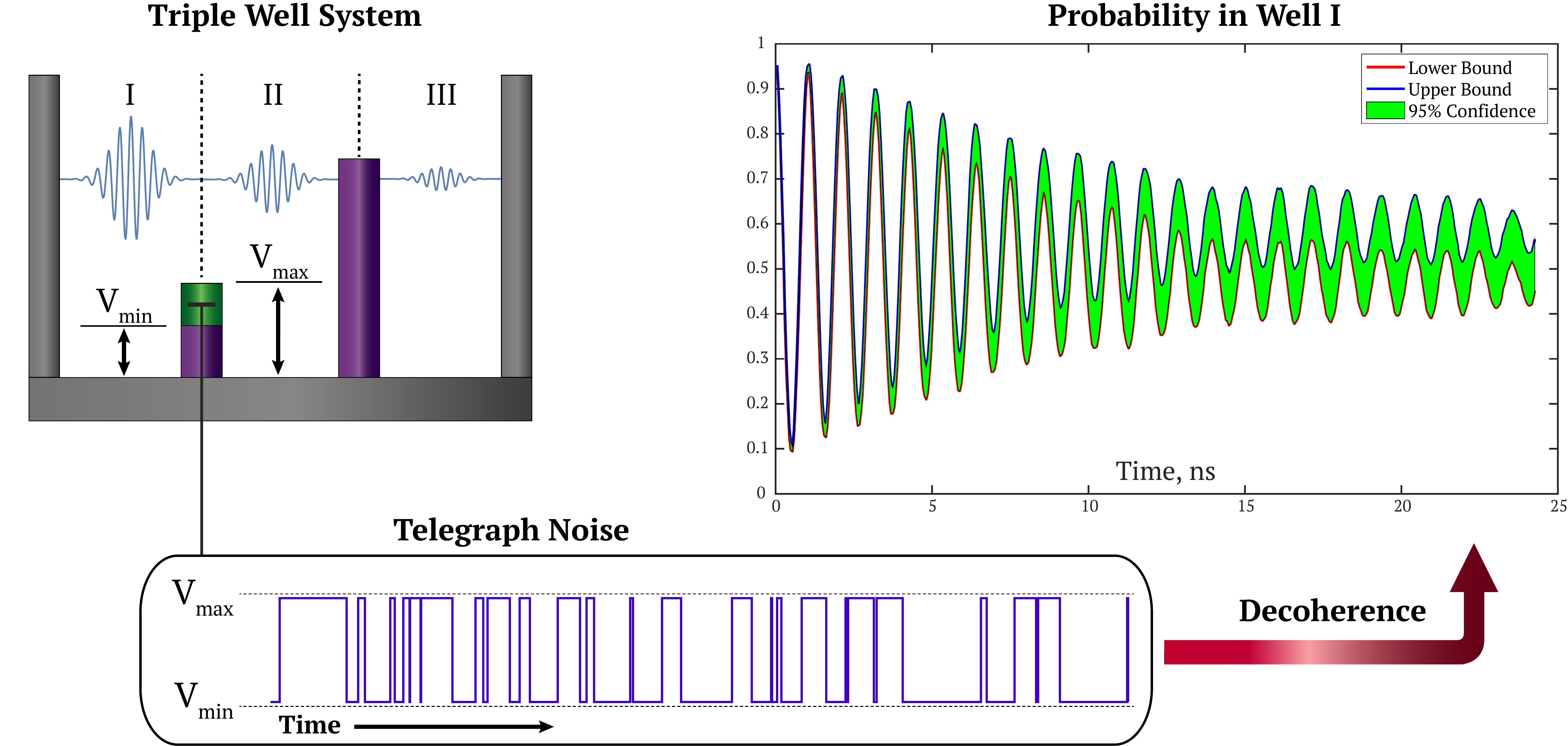}
\caption{Simulation of decoherence on example of a triple well system, where the left and right walls are $200$ units tall. The barrier that separates wells II and III is fixed and equals $64$ units. The barrier between wells I and II switched randomly between $V_{\text{min}}=4$ and $V_{\text{max}}=5$ units in accordance to the telegraph noise. The average switching time is $\approx 0.2$ ns.  } \label{fig:decoherence}
\end{figure}

\section{Discussion of Simulation Results}\label{sec:res}

Figure~\ref{fig:split_transport} demonstrates the transfer of an electron injected to  dot~1 and detected by two detectors (1 and 2) placed at the opposite edges of the structure. Two pulses of 2.6~ns duration and 60~mV amplitude are applied at imposers 1 and 2 respectively (Fig.~\ref{fig:split_transport}(a)). The heatmap (Fig.~\ref{fig:split_transport}(b)) demonstrates the PDF of the injected electron as a function of time (the horizontal axis) and coordinate along the register (the vertical axis), and its reaction to the applied pulses. The most informative is  Fig.~\ref{fig:split_transport}(c) which shows the probability of electron being in each of the dots, and the final measurement in form of a probability histogram. It is obvious that in this case it is possible to transport the electron faster, but this graph was also made to demonstrate the feasibility of Rabi oscillations with period $T_0$ between the two dots when the barrier separating the dots is lowered. Thus, to make the transport operation, we need to have the pulse duration $T_0/2 + kT_0$ with integer $k$.    

An X-rotation gate is an operation which allows one to control the probability of detecting an electron. In order to ``split'' the electron's wavefunction into two equal parts, one needs to apply a pulse of duration equal to $T_0/4 + kT_0$ (Fig.~\ref{fig:split_hadamard}(a)). Note that the second pulse applied is two times longer since we aimed at transporting a `part' of the wave function to the third well. As one can see, the shown pulses allow one to realise an X-rotation gate. However, it is evident that each operation leaves some residual wave function  in the  dots, and thus it affects the accuracy (fidelity)  of each operation. 

The presented results do not take into account decoherence effects or non-zero temperatures.  However, it would have been unfair not to mention the decoherence that results from fluctuations, which always take place in a real system.  Using a simple two-levels telegraph noise model~\cite{Abel:2008aa,Bergli:2009aa,Cai:2020aa}, we modified the algorithm described in Section~\ref{sec:split}. We run 100 simulations with the same initial conditions, where the height of the first potential barrier switches from $V_{\text{min}}$ to $V_{\text{max}}$ randomly. For each simulation, the switching time intervals were generated independently in accordance to the exponential distribution with the mean switching interval much grater than the simulation length. The corresponding results are shown in Fig.~\ref{fig:decoherence}. They demonstrate that the fluctuations of the potential energy result in the decay of Rabi oscillations. The green shadowed area corresponds to the Student's $95\%$ confidence interval.

\section{Conclusions}

In this paper, we discussed the the development of a quantum simulator for charge qubits based on FDSOI CMOS technology. We started from the description of the system and proceeded with the discussion and comparison of numerical and semi-analytical techniques to model the behaviour of a single electron in such a structure. We presented a high-level multi-particle model to simulate the evolution of the various quantum states in such a system using the tight binding approach. We demonstrated two case studies: the electron transport through multiple QDs and the construction of an X-rotation gate, and showed the effect of decoherence due to potential energy fluctuations.

\section{Acknowledgement}
A.S., D.M. and P.G. contributed equally to this work. This work was supported in part by the Science Foundation of Ireland under Grant 13/RC/2077 and under Grant 14/RP/I2921.

\bibliographystyle{splncs04}
\bibliography{biblio}

\end{document}